\documentclass[aps,prl,twocolumn,superscriptaddress]{revtex4}

\usepackage{graphicx}

\begin{document}


\title{Space-Time Clustering and Correlations of Major Earthquakes}

\author{James R. Holliday}
\email{holliday@cse.ucdavis.edu}
\affiliation{Center for Computational Science and Engineering,
             University of California, Davis}
\affiliation{Department of Physics, University of California, Davis}

\author{John B. Rundle}
\email{jbrundle@ucdavis.edu}
\affiliation{Center for Computational Science and Engineering,
             University of California, Davis}
\affiliation{Department of Physics, University of California, Davis}
\affiliation{Department of Geology, University of California, Davis}

\author{Donald L. Turcotte}
\email{turcotte@geology.ucdavis.edu}
\affiliation{Department of Geology, University of California, Davis}

\author{William Klein}
\email{klein@physics.bu.edu}
\affiliation{Department of Physics, Boston University}

\author{Kristy F. Tiampo}
\email{ktiampo@seis.es.uwo.ca}
\affiliation{Department of Earth Sciences,
             University of Western Ontario, Canada}

\date{\today}

\begin{abstract}
Earthquake occurrence in nature is thought to result from correlated
elastic stresses, leading to clustering in space and time.  We show
that occurrence of major earthquakes in California correlates with
time intervals when fluctuations in small earthquakes are suppressed
relative to the long term average.  We estimate a probability of less
than 1\% that this coincidence is due to random clustering.
\end{abstract}

\maketitle


\section{Introduction}
It is widely accepted \cite{BurridgeK67, RundleJ77, CarlsonLS94,
HelmstetterS02, MainA02, ChenBO91, Turcotte97, Sornette00,
FisherDRB97, RundleKG96, KleinRF97} that the observed earthquake
scaling laws indicate the existence of phenomena closely associated
with proximity of the system to a critical point.  More specifically,
it has been proposed that earthquake dynamics are associated either
with a second order critical point \cite{CarlsonLS94, HelmstetterS02,
MainA02, ChenBO91, Turcotte97, Sornette00, FisherDRB97} or a mean
field spinodal \cite{RundleKG96, KleinRF97} that can be understood as
a line of critical points.  Mean field theories of the Ginzburg-Landau
type have been proposed \cite{Sornette00, FisherDRB97, RundleKG96,
KleinRF97} to explain the phenomenology associated with scaling and
nucleation processes of earthquakes, which would in turn imply that a
Ginzburg criterion is applicable \cite{Goldenfeld92}.  If mean field
Ginzburg-Landau equations do describe earthquakes, the dynamics must
be operating outside the critical region, and fluctuations are
correspondingly reduced.

\subsection{To summarize our results}
We compare the performance of two probability measures that define the
locations of future earthquake occurrence: the spatially
coarse-grained seismic {\it intensity\/} and the {\it intensity
change\/}.  We show that an order parameter $\Psi_I(t)$ can be defined
based on the performance of these probability measures on a {\it
Receiver Operating Characteristic\/} (ROC) diagram and that a
generalized Ginzburg criterion $\mathcal{G}(t)$ can be established
measuring the relative importance of fluctuations in $\Psi_I(t)$.  We
find that since 1960, major earthquakes in California with magnitudes
$m\ge6$ tend to preferentially occur during intervals of time when
$\mathcal{G}(t)<1$, consistent with mean field dynamics.  Currently in
northern California, $\mathcal{G}(t)<1$.


\section{Intensity Maps and Intensity Change Maps}
\begin{figure}
\includegraphics[width=\columnwidth]{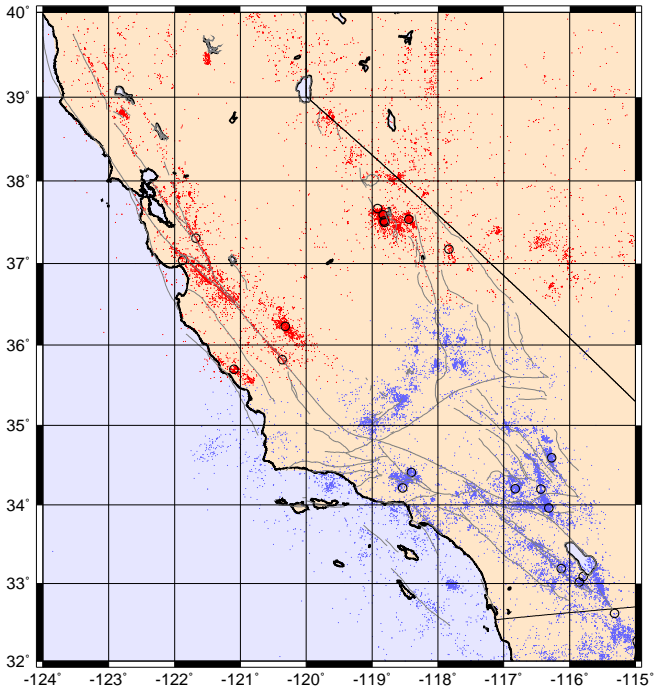}
\caption{\label{fig1}
Map of earthquake ($m\ge3$) epicenters in California from 1932 to the
present.  Circles are events with $m\ge6$ since 1960.  Red epicenters
define the area used to analyze seismicity in northern California;
blue epicenters define the area used for southern California.}
\end{figure}

The data set we use is the ANSS catalog of earthquakes
\footnote{http://www.ncedc.org/cnss/} between latitude $32^\circ$N and
$40^\circ$N and between longitudes $-124^\circ$E and $-115^\circ$E,
coarse-grained in time intervals of one day.  Only events above a
magnitude threshold $m_T\ge3$ are used to ensure catalog completeness.
Figure~\ref{fig1} shows the event locations.  We tile the region with
a spatially coarse-grained mesh of $N$ boxes, or pixels, having side
length $0.1^\circ$, about 11 km at these latitudes, approximately the
rupture length of an $m\sim6$ earthquake.  The average intensity of
activity $I(\mathbf{x}, t_0, t_2)$ is constructed by computing the
number of earthquakes $n(\mathbf{x}, t_0, t_2)$ in each coarse-grained
box centered at $\mathbf{x}$ since records began at time $t_0=1932$
until a later time $t_2$ that will be allowed to vary: $I(\mathbf{x},
t_0, t_2) = n(\mathbf{x}, t_0, t_2)$.  We then regard $P_\mu \equiv
P_\mu(\mathbf{x}, t_0, t_2) = I(\mathbf{x}, t_0, t_2) / \int
I(\mathbf{x}, t_0, t_2) \mathrm{d}\mathbf{x}$ as a probability for the
location of future events $m \ge m_T$ for times $t>t_2$.  Previous
work \cite{RundleTKM02, TiampoRMGK02a, HollidayNTRT05} indicates that
$P_\mu$ is a good predictor of locations for future large events
having $m\ge5$.

The intensity change map builds upon the intensity map by computing
the average squared change in intensity over a time interval $\Delta t
= t_2 - t_1$.  Here we use $\Delta t = 13$ years \cite{RundleTKM02,
TiampoRMGK02a}.  We compute $n(\mathbf{x}, t_b, t_1)$ and
$n(\mathbf{x}, t_b, t_2)$ for the two times $t_1$ and $t_2$, where
$t_2 > t_1$, beginning at a base time $t_b$, where $t_1>t_b>t_0$.
Computing the change in numbers of events as $\Delta n(\mathbf{x},
t_b, t_1, t_2) = n(\mathbf{x}, t_b, t_2) - n(\mathbf{x}, t_b, t_1)$,
we then define the intensity change $\Delta I(\mathbf{x}, t_1, t_2)$
by normalizing $\Delta n(\mathbf{x}, t_b, t_1, t_2)$ to have spatial
mean zero and unit variance, yielding $\Delta n'(\mathbf{x}, t_b, t_1,
t_2)$, and then averaging $\Delta n'(\mathbf{x}, t_b, t_1, t_2)$ over
all values for $t_b$ from $t_0$ to $t_1$: $\Delta I(\mathbf{x}, t_1,
t_2) = <\Delta n'(\mathbf{x}, t_b, t_1, t_2)>_{t_b}$.  The
corresponding probability is $P_\Delta \equiv P_\Delta(\mathbf{x},
t_1, t_2) = [\Delta I(\mathbf{x}, t_1, t_2)]^2 / \int [\Delta
I(\mathbf{x}, t_1, t_2)]^2 \mathrm{d}\mathbf{x}$.  Previous work
\cite{RundleTKM02, TiampoRMGK02a, HollidayNTRT05} has found that
$P_\Delta$ is also a good predictor of locations for future large
events having $m\ge5$.  $P_\Delta$ can be viewed as a probability
based upon the squared change in intensity.


\section{Binary Forecasts}
Binary forecasts are a well-known method for constructing forecasts of
future event locations and have been widely used in tornado and severe
storm forecasting \cite{HollidayNTRT05, JolliffeS03}.  We construct
binary forecasts for $m \ge m_c$ and for times $t>t_2$, where $m_c$ is
a cutoff magnitude.  In past work \cite{RundleTKM02, TiampoRMGK02a,
HollidayNTRT05} we have taken $m_c=5$, but we now remove this
restriction.  In our application, the probabilities $P_\mu \equiv
P_\mu(\mathbf{x}, t_0, t_2)$ and $P_\Delta \equiv P_\Delta(\mathbf{x},
t_1, t_2)$ are converted to binary forecasts $B_\mu \equiv B_\mu(D,
\mathbf{x}, t_0, t_2)$ and $B_\Delta \equiv B_\Delta(D, \mathbf{x},
t_1, t_2)$ by the use of a decision threshold $D$, where $D \in
[0,\max\{P_\mu\}]$ or $D \in [0,\max\{P_\Delta\}]$ respectively
\cite{HollidayNTRT05, JolliffeS03}.

\begin{figure}
\includegraphics[width=\columnwidth]{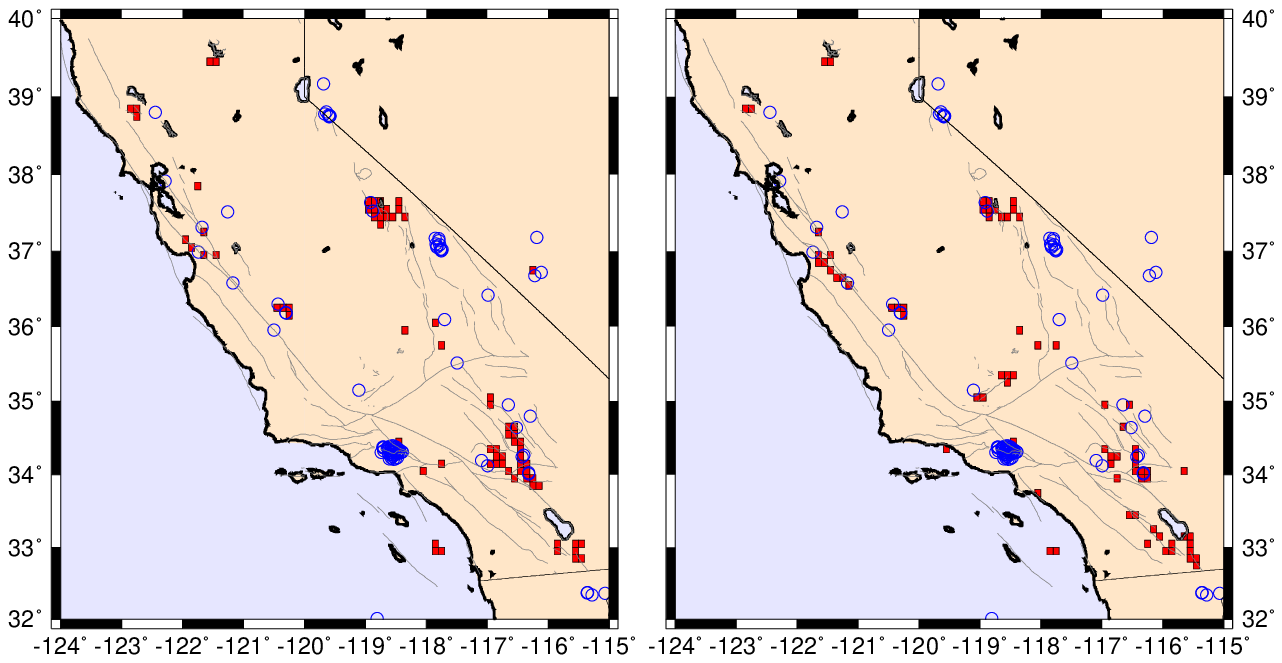}
\includegraphics[width=\columnwidth]{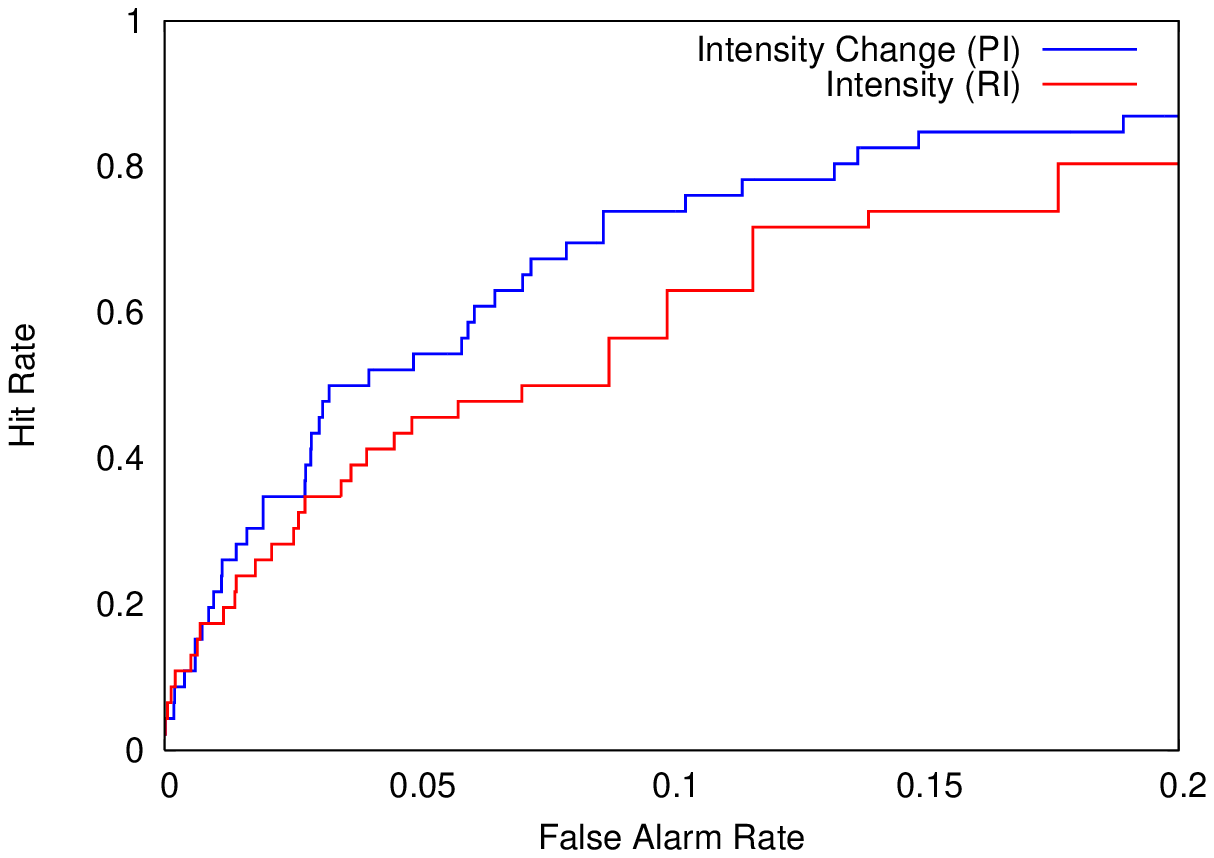}
\caption{\label{fig2}
({\bf A}) $P_\Delta$ (intensity change) map (left) and $P_\mu$
(average intensity) map (right).  A decision threshold $D$ was chosen
leading to 75 hotspots in each map.  Maps were computed for $t=$1
December 1994 with $t_2-t_1=13$ years.  ({\bf B}) ROC curves for
$P_\Delta$ and $P_\mu$ corresponding to Figure~\ref{fig1}.  Here we
have used $m_c=4$.}
\end{figure}

For a given value of $D$, we set $B_\mu=1$ where $P_\mu>D$ and
$B_\mu=0$ otherwise.  Similarly, we set $B_\Delta=1$ where
$P_\Delta>D$ and $B_\Delta=0$ otherwise.  The set of pixels
$\{\mathbf{x}_\mu(D)\}$ where $B_\mu=1$ and $\{\mathbf{x}_\Delta(D)\}$
where $B_\Delta=1$ then constitute locations where future events $m
\ge m_c$ are considered to be likely to occur.  We call these
locations {\it hotspots\/}.  The locations where $B_\mu=0$ and
$B_\Delta=0$ are sites where future events $m \ge m_c$ are unlikely to
occur.  In previous work, intensity maps and intensity change maps at
a particular value of $D$ were called {\it Relative Intensity\/} maps
and {\it Pattern Informatics\/} maps.  Examples of binary forecast
maps are shown in Figure~\ref{fig2}A.

\section{Receiver Operating Characteristic (ROC) Diagrams}
A series of $D$-dependent {\it contingency tables\/} are constructed
using the set of locations $\{\mathbf{x}_q(m_c)\}$ where the
$q=1,\ldots,Q$ large events $m \ge m_c$ are observed to actually occur
during the forecast verification period $t>t_2$. The contingency table
has 4 entries, $a \to d$, whose values are determined by some
specified rule set \cite{HollidayNTRT05, JolliffeS03}. Here we use the
following rules for given $D$ (same rules for both ``$\mu$'' and
``$\Delta$'' subscripts):
\begin{enumerate}
\item 
$a$ is the number of boxes in $\{\mathbf{x}(D)\}$ which are also in
$\{\mathbf{x}_q(m_c)\}$
\item 
$b$ is the number of boxes in $\{\mathbf{x}(D)\}$ whose location is
not in $\{\mathbf{x}_q(m_c)\}$, i.e., is in the complement to
$\{\mathbf{x}_q(m_c)\}$
\item 
$c$ is the
number of boxes in the complement to $\{\mathbf{x}(D)\}$ whose location is in
$\{\mathbf{x}_q(m_c)\}$
\item 
$d$ is the number of boxes in the complement to $\{\mathbf{x}(D)\}$
whose locations are in the complement to $\{\mathbf{x}_q(m_c)\}$
\end{enumerate}
The {\it hit rate\/} is then defined as $H = a /(a + c)$, and the {\it
false alarm rate\/} is defined as $F = b /(b + d )$.  Note that with
these definitions, $a + c = Q$, $a + b =$ number of hotspots, and $a +
b + c + d = N$.

The ROC diagram \cite{HollidayNTRT05, JolliffeS03} is a plot of the
points $\{H, F\}$ as $D$ is varied.  Examples of ROC curves
corresponding to the intensity and intensity change maps in
Figure~\ref{fig2}A are shown in Figure~\ref{fig2}B.  A perfect {\it
forecast of occurrence\/} (perfect order, no fluctuations) would
consist of two line segments, the first connecting the points $(H,F) =
(0,0)$ to $(H,F) = (1,0)$, and the second connecting $(H,F) = (1,0)$
to $(H,F) = (1,1)$.  A curve of this type can be described as maximum
possible hits ($H=1$) with minimum possible false alarms ($F=0$).
Another type of perfect forecast (perfect order, no fluctuations)
consists of two lines connecting the points $(0,0)$ to $(0,1)$ and
$(0,1)$ to $(1,1)$, a perfect {\it forecast of non-occurrence\/}.

The line $H=F$ occupies a special status, and corresponds to a
completely random forecast \cite{HollidayNTRT05, JolliffeS03} (maximum
disorder, maximum fluctuations) where the false alarm rate is the same
as the hit rate and no information is produced by the
forecast. Alternatively, we can say that the {\it marginal utility\/}
\cite{Chung94} of an additional hotspot, $\mathrm{d}H/\mathrm{d}F$,
equals unity for a random forecast.

For a given time-dependent forecast $H(F,t)$, we consider the
time-dependent {\it Pierce Skill Score\/} $H(F,t) - F$
\cite{JolliffeS03}, which measures the improvement in performance of
$H(F,t)$ relative to the random forecast $H=F$.  A {\it Pierce
function\/} $\Psi(t)$ measures the area between $H(F,t)$ and the
random forecast:
\begin{eqnarray}
\Psi(t) &=& \int_0^{F_\mathrm{max}} ( H(F,t)-F) \mathrm{d}F \\ \nonumber
        &=& A(t) - F_\mathrm{max}^2 / 2,
\end{eqnarray}
where
\begin{equation}
A(t) = \int_0^{F_\mathrm{max}} H(F,t) \mathrm{d}F
.
\end{equation}
The upper limit $F_\mathrm{max}$ on the range if integration is a
parameter whose value is set by the requirement that the marginal
utility \cite{Chung94} of the forecast of occurrence $H(F,t)$ exceeds
that of the random forecast $H=F$:
\begin{equation}
\frac{\mathrm{d}}{\mathrm{d}F}\{H(F,t)-F\} > 0
.
\end{equation}
Since $H(F,t)$ curves are monotonically increasing, $F_\mathrm{max}$
is determined as the value of $F$ for which
$\mathrm{d}H(F,t)/\mathrm{d}F=1$. For the forecasts we consider, we
find that $F_\mathrm{max}\approx0.2$, as can be seen from the examples
in Figure~\ref{fig2}B.


\section{Order Parameter and Generalized Ginzburg Criterion}
We define an order parameter as the Pierce function $\Psi_\tau(t)$
obtained using as the probability $P_\tau \equiv
P_\tau(\mathbf{x},t_1,t_2) = n(\mathbf{x},t_1,t_2) / \int
n(\mathbf{x},t_1,t_2) \mathrm{d}\mathbf{x}$, where $P_\tau$ is the
average normalized intensity of seismic activity during $t_1$ to
$t_2$. Using $P_\tau$ and the decision threshold $D$, we construct a
binary forecast $B_\tau \equiv B_\tau(D,\mathbf{x},t_1,t_2)$
. Evaluating the forecast $B_\tau$ during the time interval $t_2$ to
$t$ produces the ROC diagram $H_\tau(F,t)$.  For the case of forecasts
having positive marginal utility relative to the random forecast,
$\Psi_\tau(t)>0$.  If past seismic activity is uncorrelated with
future seismic activity, $P_\tau$ is equivalent to a random forecast,
and $\Psi_\tau(t)=0$

Corresponding to the order parameter $\Psi_\tau(t)$, we define a
function $\mathcal{G}(t)$ to indicate the relative importance of
fluctuations with respect to forecasts of occurrence.  We note that
the probability $\Psi_\Delta$ is a measure of the mean squared change
of intensity, a measure of fluctuations in seismic intensity, during
$t_1$ to $t_2$, and that the probability $P_\mu$ is a measure of the
average intensity over the entire time history ($t_0$ to $t_2$).  We
will refer to $P_\Delta$ as the ``fluctuation map'' or ``change map'',
and $P_\mu$ as the ``average map''.

Using the corresponding ROC functions we define
\begin{equation}
\mathcal{G}(t) \equiv \frac{\Psi_\Delta(t)}{\Psi_\mu(t)}
,
\end{equation}
where $\Psi_\Delta(t)$ is based upon the ROC curve computed using
$P_\Delta$, $\{H_\Delta(F,t),F\}$ and $\Psi_\mu(t)$ is based upon the
ROC curve computed using $P_\mu$, $\{H_\mu(F,t),F\}$.  We can say that
when $\mathcal{G}(t)<1$, ``fluctuations are less significant relative
to the mean'' in the sense that the fluctuation map provides a poorer
forecast than the mean map.  This statement is equivalent to the Pierce
difference function:
\begin{equation}
\Delta A(t) \equiv A_\mu(t) - A_\Delta(t) > 0
.
\end{equation}
This difference function can be considered to be a {\it generalized
Ginzburg criterion\/} \cite{Goldenfeld92}.

To examine these ideas, we compare a plot of $\mathcal{G}(t)$ with
activity of major earthquakes ($m\ge6$) in California.  We first
consider the Gutenberg-Richter frequency-magnitude relation $f = 10^a
\cdot 10^{-b m}$, where $f$ is the number of events per unit time with
magnitude larger than $m$ and $a$ and $b$ are constants.  $a$
specifies the level of activity in the region, and $b\cong1$.

To construct ROC curves, we consider $t$ to be the current time at
each time step and test the average map and change map by forecasting
locations of earthquakes during $t_2$ to $t$. We use events having $m
\ge m_T$, where $m_T$ is some threshold magnitude.  Note that $f^{-1}$
specifies a time scale for events larger than $m$: 1 event with
$m\ge6.0$ is associated on average with 10 $m\ge5.0$ events, 100
$m\ge4.0$ events, etc.  Without prior knowledge of the optimal value
for $m_T$, we average the results for a scale-invariant distribution
of 1000 $m_T\ge3.0$ events, 794 $m_T\ge3.1$ events, 631 $m_T\ge3.2$
events, $\ldots$, 10 $m_T\ge5.0$ events.  We terminate the sequence at
$m_T\ge5.0$ due to increasingly poor statistics.  To control the
number of earthquakes with $m \ge m_T$ in the {\it snapshot window\/}
($t_2$ to $t$), we determine the value of $t_2$ that most closely
produces the desired number of events within the snapshot window.  It
is possible to have fluctuations in actual number of events if the
snapshot window includes the occurrence time of a major earthquake,
when there may be many events $m \ge m_T$ in the coarse-grained time
intervals of length 1 day following the earthquake.

A central idea is that the length of the snapshot window is not fixed
in time; it is instead fixed by earthquake number at each threshold
magnitude $m_T = 3.0, 3.1, 3.2$, and so forth. Nature appears to
measure ``earthquake time'' in numbers of events, rather than in
years. ``Earthquake time'' is evidently based on stress accumulation
and release, that is, earthquake numbers, rather than in months or
years \cite{KleinRF97}.

\begin{figure}
\includegraphics[angle=270, width=0.75\columnwidth]{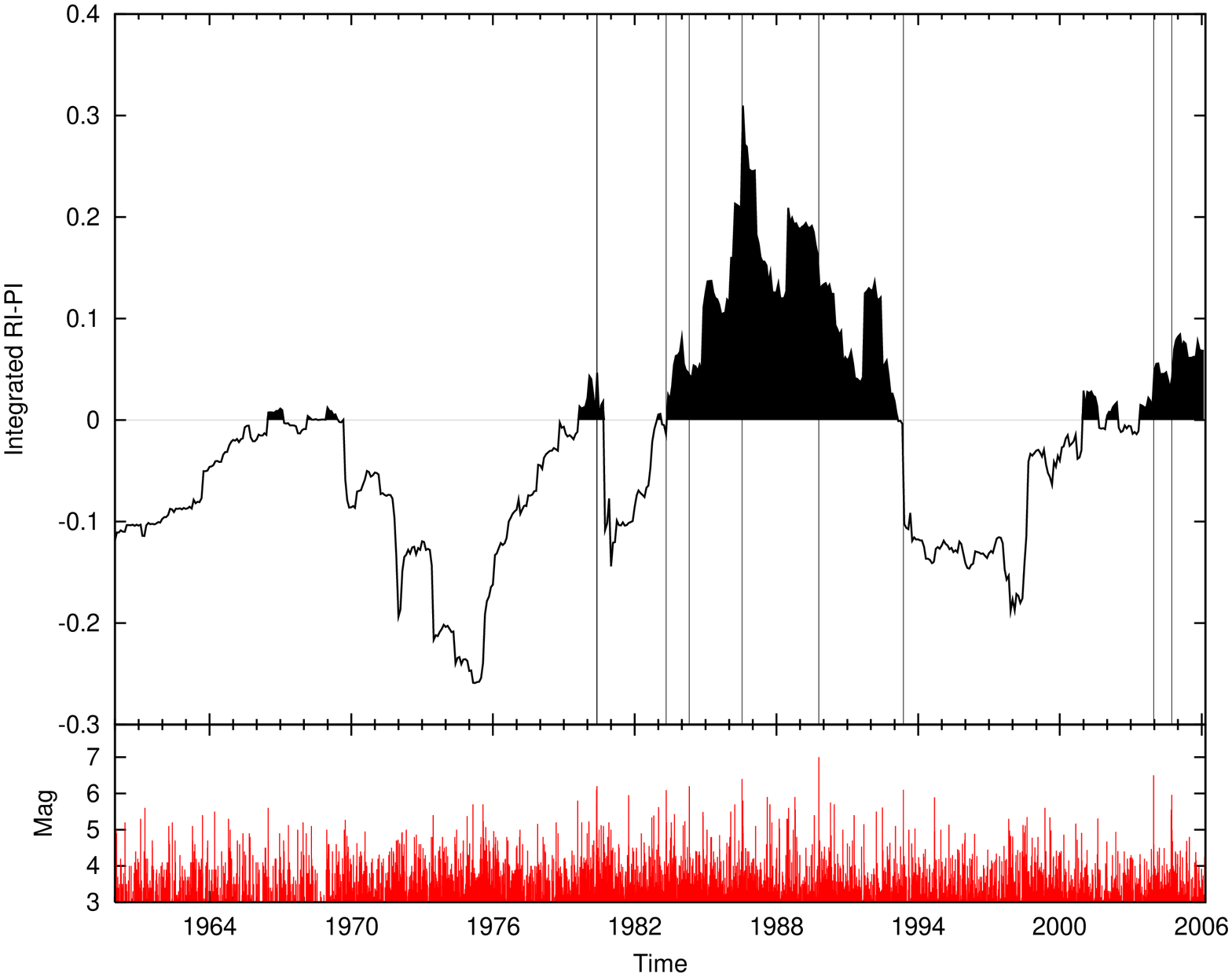}
\includegraphics[angle=270, width=0.75\columnwidth]{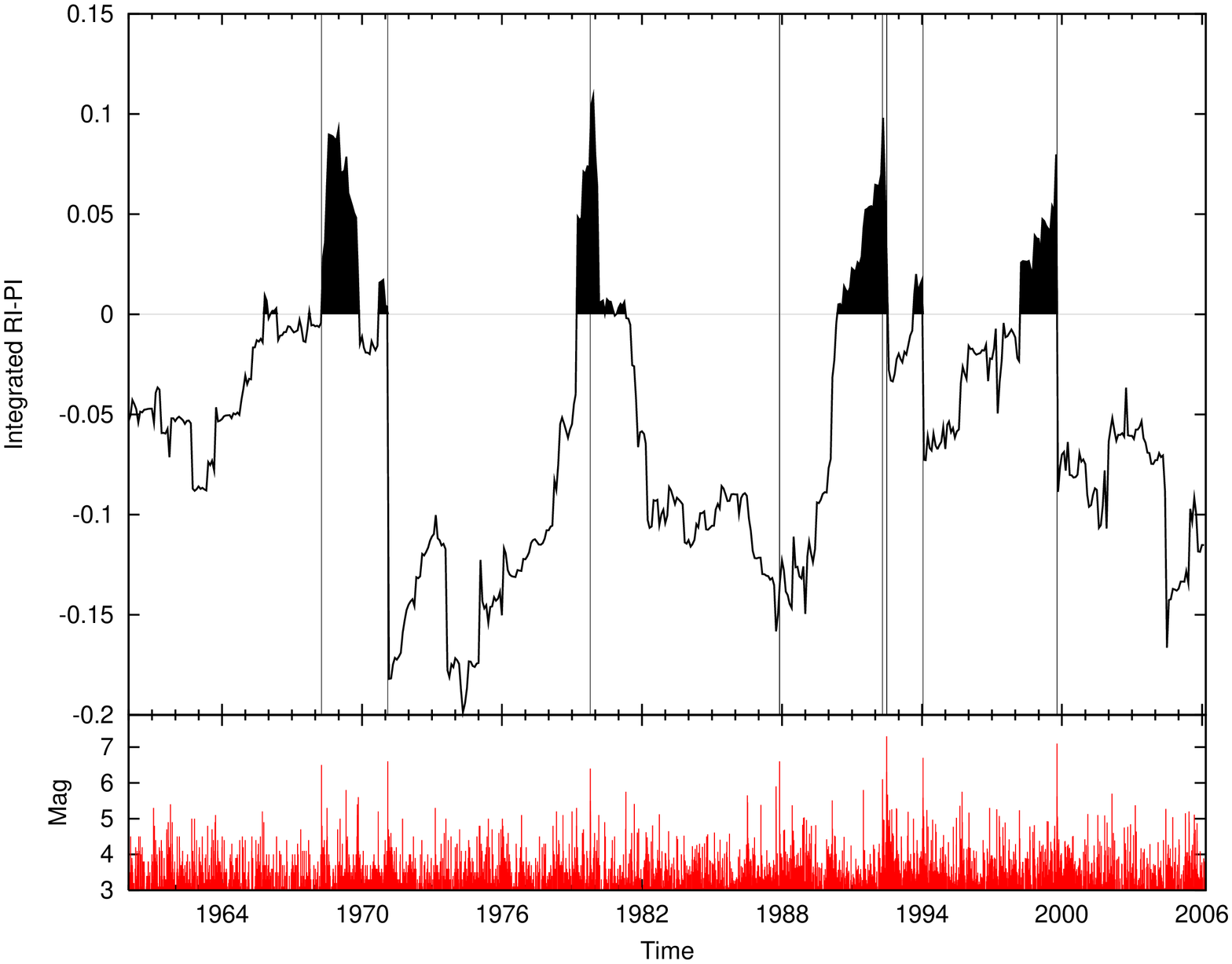}
\caption{\label{fig3}
Value of the Pierce difference function $\Delta A(t)$ (top) and
magnitude (bottom) as a function of time for events occurring on the
map area of Figure~\ref{fig1}.  Vertical black lines represent times
of major earthquakes having $m\ge6$ in the respective regions.
Differences are computed for a scale-invariant distribution of
magnitude thresholds in the snapshot window from $m_T=3.0$ to
$m_T=5.0$.  Area integration is performed for $F\in[0.0,0.2]$.  ({\bf
A}) Northern California (red epicenters in Figure~\ref{fig1}).  ({\bf
B}) Southern California (blue epicenters in Figure~\ref{fig1}).}
\end{figure}

Results are shown in Figure~\ref{fig3} for the region of California
shown in Figure~\ref{fig1}.  At top of either plot is the Pierce
difference function $\Delta A(t) = A_\mu(t) - A_\Delta(t)$, and at
bottom is earthquake magnitude plotted as a function of time from 1
January 1960 to 31 March 2006.  The vertical lines in each top panel
are the times of all events $m\ge6$ in the region during that time
interval. It can be seen from Figures~\ref{fig1} and \ref{fig3} that
there are 11 $m\ge6$ events in northern California and 10 such events
in southern California.  For both areas, these major events are
concentrated into 8 distinct ``episodes'' corresponding to 8 main
shocks.  In each plot, 7 of the 8 major episodes fall during
(``black'') time intervals where $\Delta A(t)>0$, or they either begin
or terminate such a time interval.  If a binomial probability
distribution is assumed, the chance that random clustering of these
major earthquake episodes could produce this temporal concordance can
be computed.  For Figure~\ref{fig3}A, where black time intervals
constitute 36.8\% of the total, we compute a 0.46\% chance that the
concordance is due to random clustering. For Figure~\ref{fig3}B, the
respective numbers are 19\% of the total time interval, and 0.0058\%
chance due to random clustering.  Our results support the prediction
that major earthquake episodes preferentially occur during time
intervals when fluctuations in seismic intensity, as measured by ROC
curves, are less important than the average seismic intensity.

\begin{acknowledgments}
This work has been supported by NASA Grant NGT5 to UC Davis (JRH), by
a HSERC Discovery grant (KFT), by a US Department of Energy grant to
UC Davis DE-FG03-95ER14499 (JRH and JBR), by a US Department of Energy
grant to Boston University DE-FG02-95ER14498 (WK), and through
additional funding from NSF grant ATM-0327558 (DLT).
\end{acknowledgments}




\begin{thebibliography}{17}
\expandafter\ifx\csname natexlab\endcsname\relax\def\natexlab#1{#1}\fi
\expandafter\ifx\csname bibnamefont\endcsname\relax
  \def\bibnamefont#1{#1}\fi
\expandafter\ifx\csname bibfnamefont\endcsname\relax
  \def\bibfnamefont#1{#1}\fi
\expandafter\ifx\csname citenamefont\endcsname\relax
  \def\citenamefont#1{#1}\fi
\expandafter\ifx\csname url\endcsname\relax
  \def\url#1{\texttt{#1}}\fi
\expandafter\ifx\csname urlprefix\endcsname\relax\def\urlprefix{URL }\fi
\providecommand{\bibinfo}[2]{#2}
\providecommand{\eprint}[2][]{\url{#2}}

\bibitem[{\citenamefont{Burridge and Knopoff}(1967)}]{BurridgeK67}
\bibinfo{author}{\bibfnamefont{R.}~\bibnamefont{Burridge}} \bibnamefont{and}
  \bibinfo{author}{\bibfnamefont{L.}~\bibnamefont{Knopoff}},
  \bibinfo{journal}{Bull. Seism. Soc. Am.} \textbf{\bibinfo{volume}{57}},
  \bibinfo{pages}{341} (\bibinfo{year}{1967}).

\bibitem[{\citenamefont{Rundle and Jackson}(1977)}]{RundleJ77}
\bibinfo{author}{\bibfnamefont{J.~B.} \bibnamefont{Rundle}} \bibnamefont{and}
  \bibinfo{author}{\bibfnamefont{D.~D.} \bibnamefont{Jackson}},
  \bibinfo{journal}{Bull. Seism. Soc. Am.} \textbf{\bibinfo{volume}{67}},
  \bibinfo{pages}{1363} (\bibinfo{year}{1977}).

\bibitem[{\citenamefont{Carlson et~al.}(1994)\citenamefont{Carlson, Langer, and
  Shaw}}]{CarlsonLS94}
\bibinfo{author}{\bibfnamefont{J.~M.} \bibnamefont{Carlson}},
  \bibinfo{author}{\bibfnamefont{J.~S.} \bibnamefont{Langer}},
  \bibnamefont{and} \bibinfo{author}{\bibfnamefont{B.~E.} \bibnamefont{Shaw}},
  \bibinfo{journal}{Rev. Mod. Phys.} \textbf{\bibinfo{volume}{66}},
  \bibinfo{pages}{657} (\bibinfo{year}{1994}).

\bibitem[{\citenamefont{Main and Al-Kindy}(2002)}]{MainA02}
\bibinfo{author}{\bibfnamefont{I.~G.} \bibnamefont{Main}} \bibnamefont{and}
  \bibinfo{author}{\bibfnamefont{F.~H.} \bibnamefont{Al-Kindy}},
  \bibinfo{journal}{Geophys. Res. Lett} \textbf{\bibinfo{volume}{108}},
  \bibinfo{pages}{2521} (\bibinfo{year}{2002}).

\bibitem[{\citenamefont{Chen et~al.}(1991)\citenamefont{Chen, Bak, and
  Obukhov}}]{ChenBO91}
\bibinfo{author}{\bibfnamefont{K.}~\bibnamefont{Chen}},
  \bibinfo{author}{\bibfnamefont{P.}~\bibnamefont{Bak}}, \bibnamefont{and}
  \bibinfo{author}{\bibfnamefont{S.~P.} \bibnamefont{Obukhov}},
  \bibinfo{journal}{Phys. Rev. A} \textbf{\bibinfo{volume}{43}},
  \bibinfo{pages}{625} (\bibinfo{year}{1991}).

\bibitem[{\citenamefont{Turcotte}(1997)}]{Turcotte97}
\bibinfo{author}{\bibfnamefont{D.~L.} \bibnamefont{Turcotte}},
  \emph{\bibinfo{title}{Fractals \& Chaos in Geology \& Geophysics}}
  (\bibinfo{publisher}{Cambridge University Press},
  \bibinfo{address}{Cambridge}, \bibinfo{year}{1997}), \bibinfo{edition}{2nd}
  ed.

\bibitem[{\citenamefont{Sornette}(2000)}]{Sornette00}
\bibinfo{author}{\bibfnamefont{D.}~\bibnamefont{Sornette}},
  \emph{\bibinfo{title}{Critical Phenomena in the Natural Sciences}}
  (\bibinfo{publisher}{Springer}, \bibinfo{address}{Berlin},
  \bibinfo{year}{2000}).

\bibitem[{\citenamefont{Fisher et~al.}(1997)\citenamefont{Fisher, Dahmen,
  Ramanathan, and Ben-Zion}}]{FisherDRB97}
\bibinfo{author}{\bibfnamefont{D.~S.} \bibnamefont{Fisher}},
  \bibinfo{author}{\bibfnamefont{K.}~\bibnamefont{Dahmen}},
  \bibinfo{author}{\bibfnamefont{S.}~\bibnamefont{Ramanathan}},
  \bibnamefont{and} \bibinfo{author}{\bibfnamefont{Y.}~\bibnamefont{Ben-Zion}},
  \bibinfo{journal}{Phys. Rev. Lett.} \textbf{\bibinfo{volume}{78}},
  \bibinfo{pages}{4885} (\bibinfo{year}{1997}).

\bibitem[{\citenamefont{Rundle et~al.}(1996)\citenamefont{Rundle, Klein, and
  Gross}}]{RundleKG96}
\bibinfo{author}{\bibfnamefont{J.~B.} \bibnamefont{Rundle}},
  \bibinfo{author}{\bibfnamefont{W.}~\bibnamefont{Klein}}, \bibnamefont{and}
  \bibinfo{author}{\bibfnamefont{S.~J.} \bibnamefont{Gross}},
  \bibinfo{journal}{Phys. Rev. Lett.} \textbf{\bibinfo{volume}{76}},
  \bibinfo{pages}{4285} (\bibinfo{year}{1996}).

\bibitem[{\citenamefont{Klein et~al.}(1997)\citenamefont{Klein, Rundle, and
  Ferguson}}]{KleinRF97}
\bibinfo{author}{\bibfnamefont{W.}~\bibnamefont{Klein}},
  \bibinfo{author}{\bibfnamefont{J.~B.} \bibnamefont{Rundle}},
  \bibnamefont{and} \bibinfo{author}{\bibfnamefont{C.~D.}
  \bibnamefont{Ferguson}}, \bibinfo{journal}{Phys. Rev. Lett.}
  \textbf{\bibinfo{volume}{78}}, \bibinfo{pages}{3793} (\bibinfo{year}{1997}).

\bibitem[{\citenamefont{Helmstetter and Sornette}(2002)}]{HelmstetterS02}
\bibinfo{author}{\bibfnamefont{A.}~\bibnamefont{Helmstetter}} \bibnamefont{and}
  \bibinfo{author}{\bibfnamefont{D.}~\bibnamefont{Sornette}},
  \bibinfo{journal}{J. Geophys. Res.} \textbf{\bibinfo{volume}{107}},
  \bibinfo{pages}{2237} (\bibinfo{year}{2002}).

\bibitem[{\citenamefont{Goldenfeld}(1992)}]{Goldenfeld92}
\bibinfo{author}{\bibfnamefont{N.}~\bibnamefont{Goldenfeld}},
  \emph{\bibinfo{title}{Lectures on Phase Transitions and the Renormalization
  Group}} (\bibinfo{publisher}{Addison Wesley}, \bibinfo{address}{Reading, MA},
  \bibinfo{year}{1992}).

\bibitem[{\citenamefont{Rundle et~al.}(2002)\citenamefont{Rundle, Tiampo,
  Klein, and Martins}}]{RundleTKM02}
\bibinfo{author}{\bibfnamefont{J.~B.} \bibnamefont{Rundle}},
  \bibinfo{author}{\bibfnamefont{K.~F.} \bibnamefont{Tiampo}},
  \bibinfo{author}{\bibfnamefont{W.}~\bibnamefont{Klein}}, \bibnamefont{and}
  \bibinfo{author}{\bibfnamefont{J.~S.~S.} \bibnamefont{Martins}},
  \bibinfo{journal}{Proc. Natl. Acad. Sci. U.~S.~A.}
  \textbf{\bibinfo{volume}{99}}, \bibinfo{pages}{2514} (\bibinfo{year}{2002}).

\bibitem[{\citenamefont{Tiampo et~al.}(2002)\citenamefont{Tiampo, Rundle,
  McGinnis, Gross, and Klein}}]{TiampoRMGK02a}
\bibinfo{author}{\bibfnamefont{K.~F.} \bibnamefont{Tiampo}},
  \bibinfo{author}{\bibfnamefont{J.~B.} \bibnamefont{Rundle}},
  \bibinfo{author}{\bibfnamefont{S.}~\bibnamefont{McGinnis}},
  \bibinfo{author}{\bibfnamefont{S.~J.} \bibnamefont{Gross}}, \bibnamefont{and}
  \bibinfo{author}{\bibfnamefont{W.}~\bibnamefont{Klein}}, \bibinfo{journal}{J.
  Geophys. Res.} \textbf{\bibinfo{volume}{107}}, \bibinfo{pages}{2354}
  (\bibinfo{year}{2002}).

\bibitem[{\citenamefont{Holliday et~al.}(2005)\citenamefont{Holliday, Nanjo,
  Tiampo, Rundle, and Turcotte}}]{HollidayNTRT05}
\bibinfo{author}{\bibfnamefont{J.~R.} \bibnamefont{Holliday}},
  \bibinfo{author}{\bibfnamefont{K.~Z.} \bibnamefont{Nanjo}},
  \bibinfo{author}{\bibfnamefont{K.~F.} \bibnamefont{Tiampo}},
  \bibinfo{author}{\bibfnamefont{J.~B.} \bibnamefont{Rundle}},
  \bibnamefont{and} \bibinfo{author}{\bibfnamefont{D.~L.}
  \bibnamefont{Turcotte}}, \bibinfo{journal}{Nonlinear Processes in Geophysics}
  \textbf{\bibinfo{volume}{12}}, \bibinfo{pages}{965} (\bibinfo{year}{2005}).

\bibitem[{\citenamefont{Jolliffe and Stephenson}(2003)}]{JolliffeS03}
\bibinfo{author}{\bibfnamefont{I.~T.} \bibnamefont{Jolliffe}} \bibnamefont{and}
  \bibinfo{author}{\bibfnamefont{D.~B.} \bibnamefont{Stephenson}},
  \emph{\bibinfo{title}{Forecast Verification}} (\bibinfo{publisher}{John
  Wiley}, \bibinfo{address}{Chichester}, \bibinfo{year}{2003}).

\bibitem[{\citenamefont{Chung}(1994)}]{Chung94}
\bibinfo{author}{\bibfnamefont{J.~W.} \bibnamefont{Chung}},
  \emph{\bibinfo{title}{Utility and Production Functions}}
  (\bibinfo{publisher}{Blackwell}, \bibinfo{address}{Oxford},
  \bibinfo{year}{1994}).

\end{thebibliography}
\end{document}